\documentclass{article}

\usepackage{graphicx} 
\usepackage{subfigure} 
\usepackage{placeins} 
\usepackage{natbib}
\usepackage{comment}
\usepackage{enumerate}
\usepackage{dblfloatfix}
\usepackage{algorithm}
\usepackage{algorithmic}

\usepackage{hyperref}

\usepackage{amsmath}
\usepackage{amssymb}
\usepackage{amsthm}
\usepackage{upgreek}
\theoremstyle{plain}
\newtheorem{theorem}{Theorem}
\newtheorem{fact}{Fact}
\newtheorem{proposition}{Proposition}

\newtheorem{lemma}{Lemma}

\theoremstyle{definition}
\newtheorem{definition}{Definition}

\theoremstyle{remark}

\newcommand{\R}{\mathbb{R}}

\renewcommand{\hat}[1]{\widehat{#1}}

\newcommand{\A}{\mathcal{A}}

\renewcommand{\P}{\mathbb{P}}
\newcommand{\E}{\mathbb{E}}

\newcommand{\e}{\epsilon}
\newcommand{\ve}{\varepsilon}

\newcommand{\tr}{\operatorname{tr}}

\newcommand{\var}{\operatorname{var}}

\newcommand{\ttop}{^{\top}}

\newcommand{\op}{_{\text{op}}}

\newcommand{\ts}{\textstyle}

\newcommand{\err}{\text{err}}
\newcommand{\argmin}{\text{argmin}}
\newcommand{\ccirc}{^{\circ}}
\newcommand{\med}{\text{median}}

\newcommand{\mnorm}[1]{\left\vert\kern-1.5pt\left\vert\kern-1.5pt\left\vert #1\right\vert\kern-1.5pt\right\vert\kern-1.5pt\right\vert}

\newenvironment{myenumerate}{%
  \edef\backupindent{\the\parindent}%
  \itemize%
  \setlength{\parindent}{\backupindent}%
}{\enditemize}

\begin{document} 

\begin{center}

{\bf{\LARGE{Estimating Unknown Sparsity in Compressed Sensing}}}

\vspace*{.3in}

{\large{Miles E. Lopes}}\\

{\large{\texttt{mlopes@stat.berkeley.edu}}}

{\large{
\vspace*{.2in}
Department of Statistics\\
University of California, Berkeley

}}
\end{center}
~\\
\begin{abstract}
In the theory of compressed sensing (CS), the sparsity $\|x\|_0$ of the unknown signal \mbox{$x\in\R^p$} is commonly assumed to be a known parameter. However, it is typically unknown in practice. Due to the fact that many aspects of  CS depend on knowing $\|x\|_0$, it is important to estimate this parameter in a data-driven way. A second practical concern is that $\|x\|_0$ is a highly unstable function of $x$. In particular, for real signals with entries not exactly equal to 0, the value $\|x\|_0=p$ is not a useful description of the effective number of coordinates. In this paper, we propose to estimate a stable measure of sparsity $s(x):=\|x\|_1^2/\|x\|_2^2$, which is a sharp lower bound on $\|x\|_0$. Our estimation procedure uses only a small number of linear measurements, does not rely on any sparsity assumptions, and requires very little computation. A confidence interval for $s(x)$ is  provided, and its width is shown to have no dependence on the signal dimension $p$.  Moreover, this result extends naturally to the matrix recovery setting, where a soft version of matrix rank can be estimated with analogous guarantees. Finally, we show that the use of randomized measurements is essential to estimating $s(x)$. This is accomplished by proving that the minimax risk for estimating $s(x)$ with deterministic measurements is large when $n\ll p$.

\end{abstract}

\section{Introduction}
The central problem of compressed sensing (CS) is to estimate an unknown 
signal $x\in \R^p$ from $n$ linear measurements $y=(y_1,\dots,y_n)$ given by
\begin{equation}\label{setup1}
y=Ax+\e,
\end{equation}
where $A\in\R^{n\times p}$ is a user-specified measurement matrix, $\e\in\R^n$ is a random noise vector, and $n$ is much smaller than the signal dimension $p$. During the last several years, the theory of CS has drawn widespread attention to the fact that this seemingly ill-posed problem can be solved reliably when $x$ is sparse --- 
in the sense that the parameter \mbox{$\|x\|_0:=\text{card}\{j : x_j \neq 0\}$} is much less than $p$.
For instance, if $n$ is approximately \mbox{$\|x\|_0\log(p/|x\|_0)$}, then accurate recovery can be achieved with high probability when $A$ is drawn from a Gaussian ensemble~\cite{,donoho2006,candes2006}. Along these lines, the value of the parameter $\|x\|_0$ is commonly assumed to be known in the analysis of recovery algorithms --- even though it is typically \emph{unknown} in practice. Due to the fundamental role that sparsity plays in CS, this issue has been recognized as a significant gap between theory and practice by several authors~\cite{wardCV,eldarStein,willsky}. Nevertheless, the literature has been relatively quiet about the problems of estimating this parameter and quantifying its uncertainty.
\subsection{Motivations and the role of sparsity}\label{sec:role}
 At a conceptual level, the problem of estimating $\|x\|_0$ is quite different from the more well-studied problems of estimating the full signal $x$ or its support set \mbox{$S:=\{j: x_j\neq 0\}$}.
 The difference arises from sparsity assumptions. On one hand, a procedure for estimating $\|x\|_0$ should make very few assumptions about sparsity (if any). On the other hand, methods for estimating $x$ or $S$ often assume that a sparsity level is given, and then   
\emph{impose} this value on the solution $\hat{x}$ or $\hat{S}$.
Consequently, a simple plug-in estimate of $\|x\|_0$, such as $\|\hat{x}\|_0$ or $\text{card}(\hat{S})$, may fail when the sparsity assumptions underlying $\hat{x}$ or $\hat{S}$ are invalid.

To emphasize that there are many aspects of CS that depend on knowing $\|x\|_0$, we provide several examples below. Our main point here is that a method for estimating $\|x\|_0$ is valuable because it can help to address a broad range of issues.

\begin{myenumerate}

 \item {\bf{Modeling assumptions.}} One of the core modeling assumptions invoked in applications of CS is that the signal of interest has a sparse representation. Likewise, the problem of checking whether or not this assumption is supported by data has been an active research topic, particularly in in areas of face recognition and image classification~\cite{rigamonti2011sparse,shi2011face}. In this type of situation, an estimate $\hat{\|x\|}_0$ that does not rely on any sparsity assumptions is a natural device for validating the use of sparse representations.

\item {\bf{The number of measurements.}} 
If the choice of $n$ is too small compared to the ``critical'' number $n^*(x):=\|x\|_0\log(p/\|x\|_0)$, then there are known information-theoretic barriers to the accurate reconstruction of $x$~\cite{arias2011fundamental}.
At the same time, if $n$ is chosen to be much larger than $n^*(x)$, then the measurement process is wasteful (since there are known algorithms that can reliably recover $x$ with approximately $n^*(x)$ measurements~\cite{eldarintro}). 

To deal with the selection of $n$, a sparsity estimate $\hat{\|x\|}_0$ may be used in two different ways,  depending on whether measurements are collected sequentially, or in a single batch. In the sequential case, an estimate of $\|x\|_0$ can be computed from a set of ``preliminary'' measurements, and then the estimated value $\hat{\|x\|}_0$ determines how many additional measurements should be collected to recover the full signal. Also, it is not always necessary to take additional measurements, since the preliminary set may be re-used to compute $\hat{x}$ (as discussed in Section~\ref{sec:sims}).  Alternatively, if all of the measurements must be taken in one batch, the value $\widehat{\|x\|}_0$ can be used to certify whether or not enough measurements were actually taken.

\item {\bf{The measurement matrix.}} Two of the most well-known design characteristics of the matrix $A$ are defined explicitly in terms of sparsity. These are the  \emph{restricted isometry property of order $k$} (RIP-$k$), and the \emph{restricted null-space property of order $k$} (NSP-$k$), where $k$ is a presumed upper bound on the sparsity level of the true signal.
Since many recovery guarantees are closely tied to RIP-$k$
 and NSP-$k$, a growing body of work has been devoted to certifying whether or not a given matrix satisfies these properties~\cite{daspremont,nemirovski, tangnehorai}.
When $k$ is treated as given, this problem is already computationally difficult. Yet, when the sparsity of $x$ is unknown, we must also remember that such a ``certificate'' is not meaningful unless we can check that $k$ is consistent with the true signal.

 \item {\bf{Recovery algorithms.}} When recovery algorithms are implemented, the sparsity level of $x$ is often treated as a tuning parameter. For example, if $k$ is a conjectured bound on $\|x\|_0$, then the Orthogonal Matching Pursuit algorithm (OMP) is typically initialized to run for $k$ iterations. A second example is the Lasso algorithm, which computes the solution \mbox{$\hat{x}\in\argmin\{ \|y-Av\|_2^2+\lambda\|v\|_1 : v\in\R^p\}$,} for  some choice of $\lambda\geq 0$. The sparsity of $\hat{x}$ is determined by the size of $\lambda$, and in order to select the appropriate value, a family of solutions is examined over a range of $\lambda$ values~\cite{lars}. 
 In the case of either OMP or Lasso, a sparsity estimate $\hat{\|x\|}_0$ would reduce computation by restricting the possible choices of $\lambda$ or $k$, and it would also ensure that the chosen values conform to the true signal.
 \end{myenumerate}
%
\begin{figure*}[!b]
\centering
{\includegraphics[angle=0,
  width=.3\linewidth]{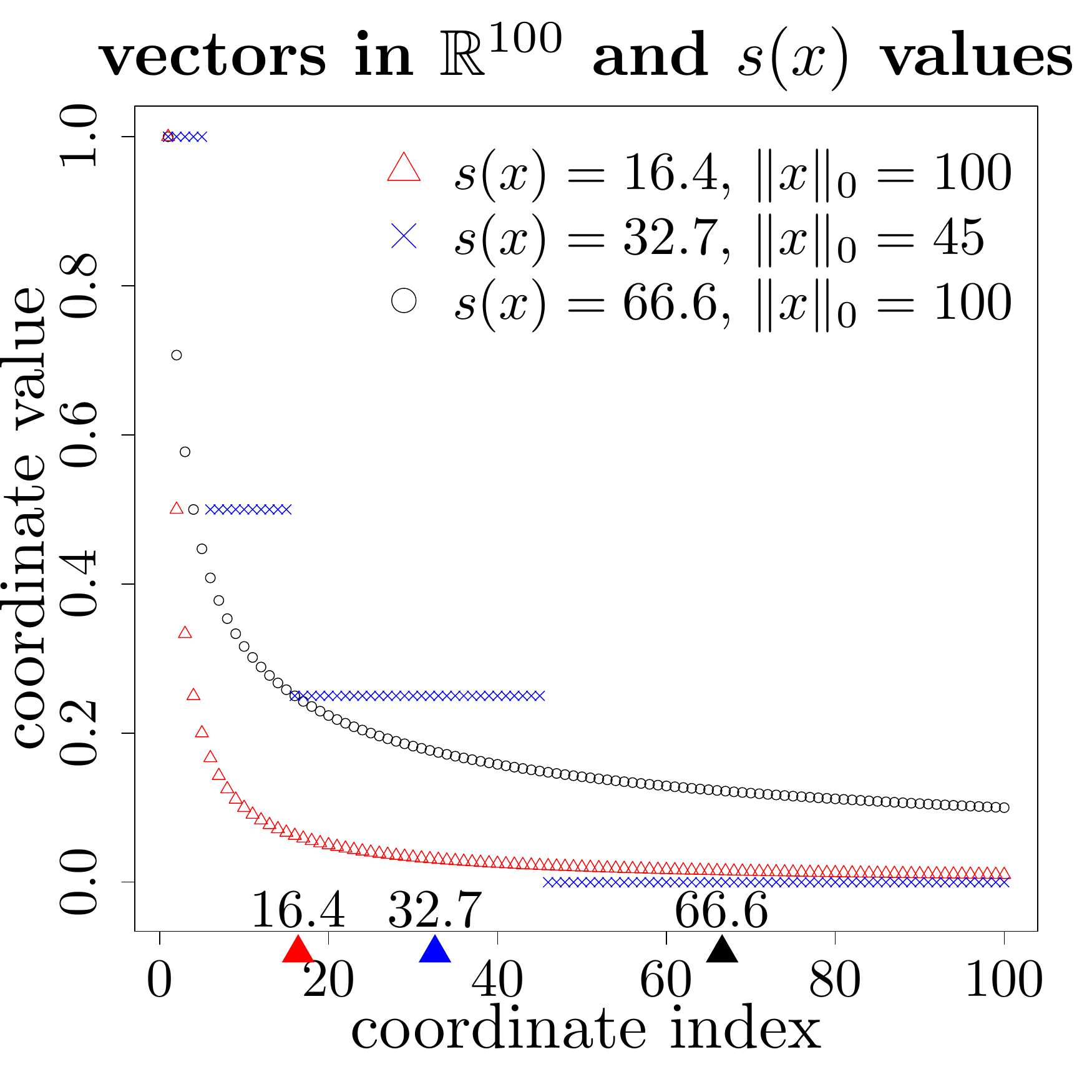}}
{\includegraphics[angle=0,
  width=.3\linewidth]{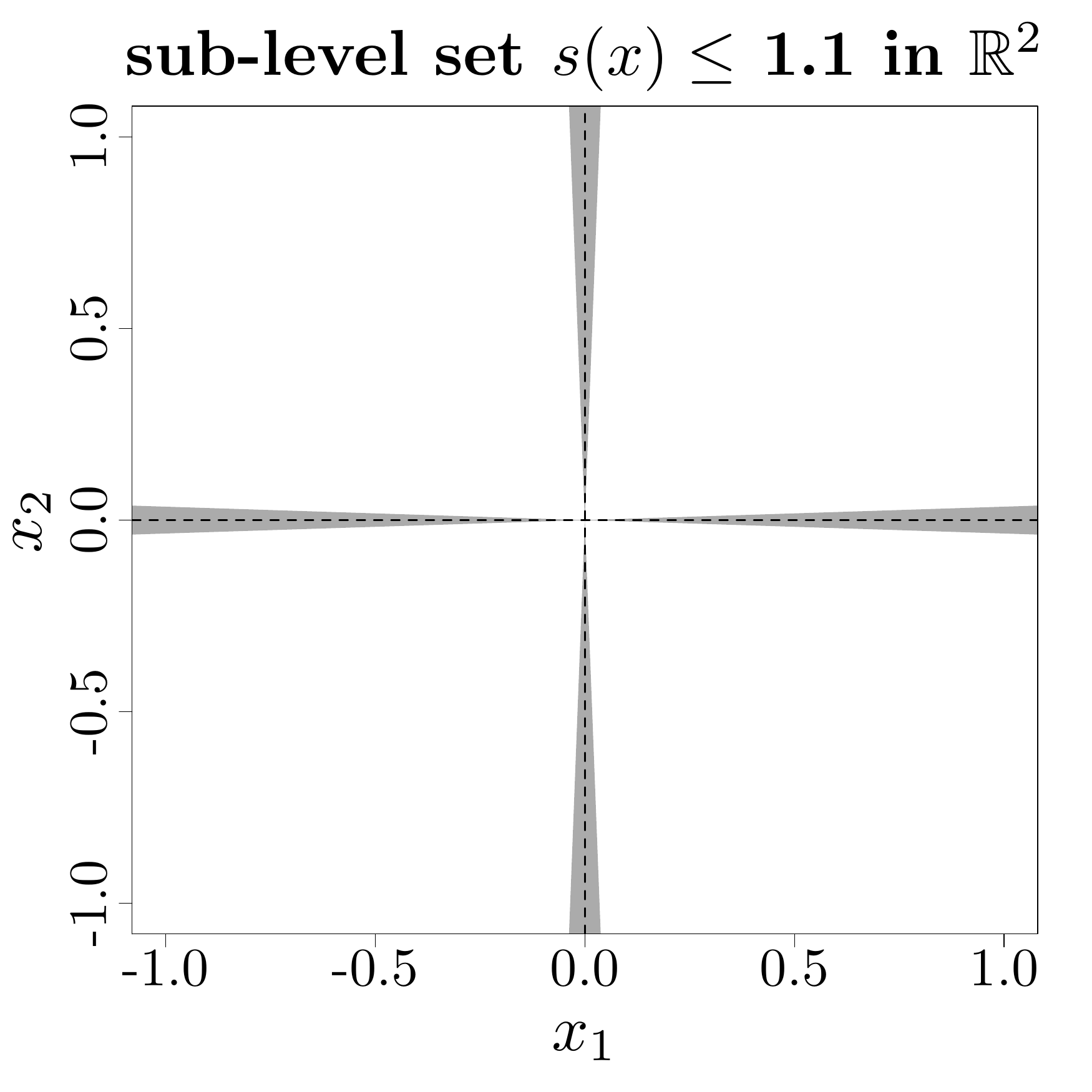}}
{\includegraphics[angle=0,
  width=.3\linewidth]{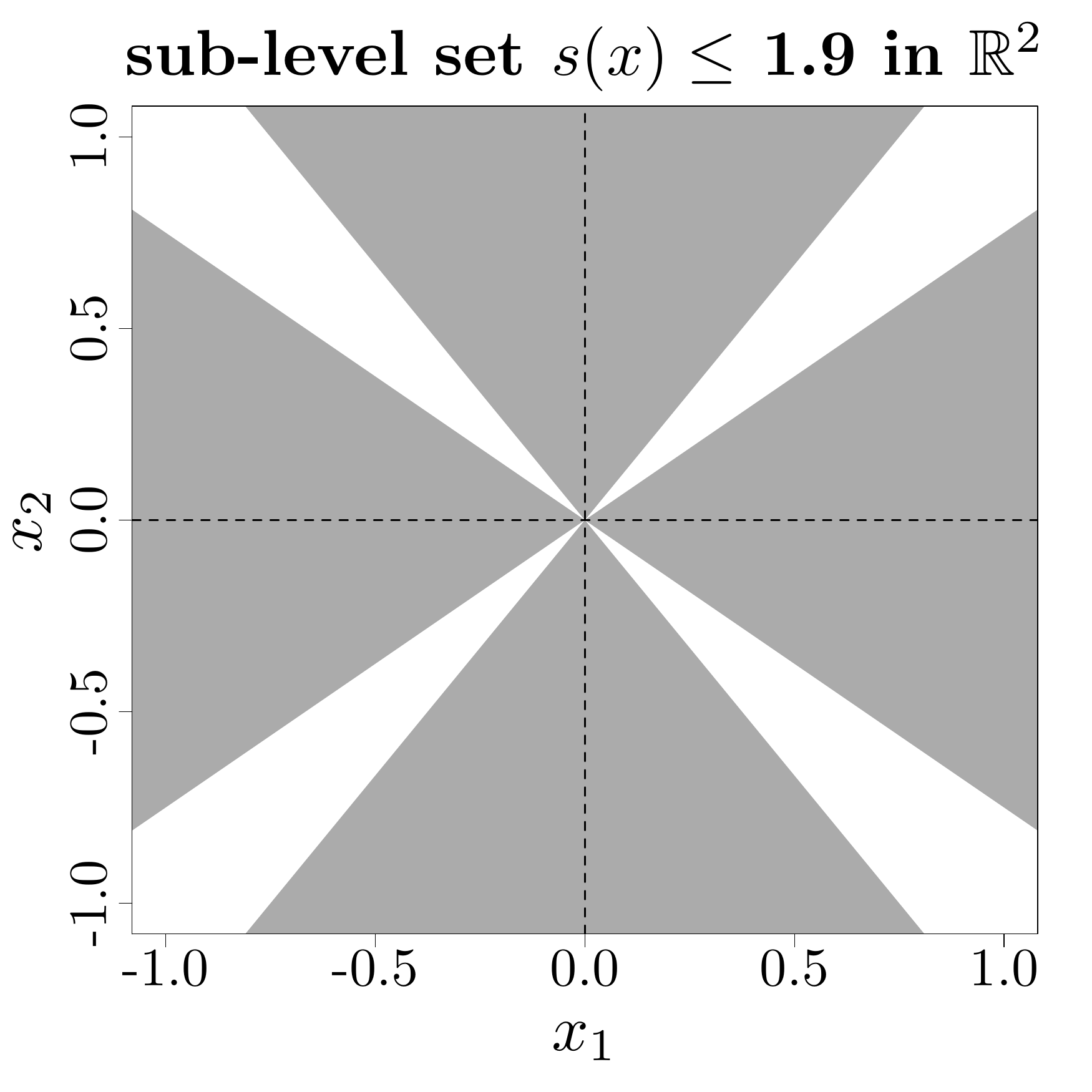}}

\caption{Characteristics of $s(x)$. Left panel: Three vectors (red, blue, black) in $\R^{100}$ have been plotted with their coordinates in order of decreasing size (maximum entry normalized to 1). Two of the vectors have power-law decay profiles, and one is a dyadic vector with exactly 45 positive coordinates (red: $x_i \propto i^{-1}$, blue: dyadic, black: $x_i\propto \ts i^{-1/2}$).  Color-coded triangles on the bottom axis indicate that the $s(x)$ value represents the ``effective'' number of coordinates. 
}
\label{fig:sparse}
\end{figure*}

\subsection{An alternative measure of sparsity}
Despite the important theoretical role of the parameter $\|x\|_0$ in many aspects of CS, it has the practical drawback of being a highly unstable function of $x$. In particular, for real signals $x\in\R^p$ whose entries are not exactly equal to 0, the value $\|x\|_0=p$ is not a useful description of the effective number of coordinates. 

In order to estimate sparsity in a way that accounts for the instability of $\|x\|_0$, it is desirable to replace the $\ell_0$ norm with a ``soft'' version. More precisely, we would like to identify a function of $x$ that can be interpreted like $\|x\|_0$, but remains stable under small perturbations of $x$. A natural quantity that serves this purpose is the \emph{numerical sparsity}
\begin{equation}\label{sdef}
s(x):=\frac{\|x\|_1^2}{\|x\|_2^2},
\end{equation}
which always satisfies $1\leq s(x)\leq p$ for any non-zero $x$. Although the ratio $\|x\|_1^2/\|x\|_2^2$ appears sporadically in different areas~\cite{tangnehorai,measures,hoyer,lopes}, it does not seem to be well known as a sparsity measure in CS. Our terminology derives from the so-called \emph{numerical rank} $\mnorm{X}_F^2/\mnorm{X}\op^2$ of a matrix $X$, which is a soft version of the rank function, coined by~\cite{vershyninnumerical}.

A key property of $s(x)$ is that it is a sharp lower bound on $\|x\|_0$ for all non-zero $x$,
\begin{equation}\label{slowerbound}
 \ \ \ s(x)\leq \|x\|_0, 
\end{equation}
which follows from applying the Cauchy-Schwarz inequality to the relation $\|x\|_1=\langle x,\text{sgn}(x)\rangle$. (Equality in~\eqref{slowerbound} is attained iff the non-zero coordinates of $x$ are equal in magnitude.) We also note that this inequality is invariant to scaling of $x$, since $s(x)$ and $\|x\|_0$ are individually scale invariant. In the opposite direction, it is easy to see that the only continuous upper bound  on $\|x\|_0$ is the trivial one: If a continuous function $f$ satisfies \mbox{$\|x\|_0 \leq f(x)\leq p$} for all $x$ in some open subset of $\R^p$, then $f$ must be identically equal to $p$. Therefore, we must be content with a continuous lower bound.

The fact that $s(x)$ is a sensible measure of sparsity for non-idealized signals is illustrated in Figure 1. In essence, if $x$ has $k$ large coordinates and $p-k$ small coordinates, then $s(x)\approx k$, whereas $\|x\|_0=p$. In the left panel, the sorted coordinates of three different vectors in $\R^{100}$ are plotted. The value of $s(x)$ for each vector is marked with a triangle on the x-axis, which shows that $s(x)$ adapts well to the decay profile. This idea can be seen in a more geometric way in the middle and right panels, which plot the  the sub-level sets $\mathcal{S}_c:=\{x \in \R^p : s(x)\leq c\}$ with $c\in [1,p]$. When $c\approx 1$, the vectors in $\mathcal{S}_c$ are closely aligned with the coordinate axes, and hence contain one effective coordinate. As $c\uparrow p$, the set $\mathcal{S}_c$ expands to include less sparse vectors until $\mathcal{S}_p=\R^p$.
\subsection{Related work.}
Some of the challenges described in Section~\ref{sec:role} can be approached with the general tools of cross-validation (CV) and empirical risk minimization (ERM). This approach has been used to select various parameters, such as the number of measurements $n$~\cite{willsky,wardCV}, the number of OMP iterations $k$~\cite{wardCV}, or the Lasso regularization parameter $\lambda$~\cite{eldarStein}.  At a high level, these methods consider a collection of (say $m$) solutions $\hat{x}^{(1)},\dots,\hat{x}^{(m)}$ obtained from different values $\theta_1,\dots,\theta_m$ of some tuning parameter of interest. For each solution, an empirical error estimate $\hat{\err}(\hat{x}^{(j)})$ is computed, and the value $\theta_{j^*}$ corresponding to the smallest $\hat{\err}(\hat{x}^{(j)})$ is chosen. 

Although methods based on CV/ERM share common motivations with our work here, these methods differ from our approach in several ways.  In particular, the problem of estimating a soft measure of sparsity, such as $s(x)$, has not been considered from that angle. Also, the cited methods do not give any theoretical guarantees to ensure that the estimated sparsity level is close to the true one. Note that even if an estimate $\hat{x}$ has small risk $\|\hat{x}-x\|_2$, it is not necessary for $\|\hat{x}\|_0$ to be close to $\|x\|_0$. This point is relevant in contexts where one is  interested in identifying a set of important variables or interpreting features.

From a computational point view, the CV/ERM approaches can also be costly --- since $\hat{x}^{(j)}$ must often be computed from a separate optimization problem for  for each choice of the tuning parameter. 
By contrast, our method for estimating $s(x)$ requires no optimization and can be computed easily from just a small set of preliminary measurements.

\subsection{Our contributions.} The primary contributions of this paper consist in
identifying a stable measure of sparsity that is relevant to CS, and proposing an efficient estimator with  with provable guarantees. Secondly, we are not aware of any other papers that have identified a distinction between random and deterministic measurements with regard to estimating unknown sparsity (as in Section~\ref{sec:neg}).

The remainder of the paper is organized as follows. In Section~\ref{sec:recov}, we show that a principled choice of $n$ can be made if $s(x)$ is known. This is accomplished by formulating a recovery condition for the Basis Pursuit algorithm directly in terms of $s(x)$. Next, in Section~\ref{sec:est}, we propose an estimator $\hat{s}(x)$, and derive a dimension-free confidence interval for $s(x)$. The procedure is also shown to extend to the problem of estimating a soft measure of rank for matrix-valued signals.  In Section~\ref{sec:neg} we show that the use of randomized measurements is essential to estimating $s(x)$ in a minimax sense. Finally, we present simulations in Section~\ref{sec:sims} to validate the consequences of our theoretical results. We defer all of our proofs to the appendix.

\paragraph{Notation.}
We define $\|x\|_q^q:=\sum_{j=1}^p |x_j|^q$ for any $q>0$ and $x\in \R^p$, which only corresponds to a genuine norm for $q\geq 1$.
For sequences of numbers $a_n$ and $b_n$, we write $a_n\lesssim b_n$ or $a_n=\mathcal{O}(b_n)$ if there is an absolute constant $c>0$ such that $a_n \leq c b_n$ for all large $n$. If $a_n/b_n\to 0$, we write $a_n=o(b_n)$.
For a matrix $M$, we define 
the Frobenius norm $\|M\|_F=\sqrt{\sum_{i,j} M_{ij}^2}$, the 
matrix $\ell_1$-norm $\|M\|_1 = \sum_{i,j} |M_{ij}|$.
Finally, for two matrices $A$ and $B$ 
of the same size, we define the inner product 
$\langle A,B \rangle := \tr(A\ttop B)$.  
%
%
%
%
%
%
 %
 %
%
%
%

%
\section{Recovery conditions in terms of $s(x)$}\label{sec:recov}
The purpose of this section is to present a simple proposition that links $s(x)$ with recovery conditions for the Basis Pursuit algorithm (BP). This is an important motivation for studying $s(x)$, since it implies that if $s(x)$ can be estimated well, then $n$ can be chosen appropriately.

In order to explain the connection between $s(x)$ and recovery, we first recall a standard result~\cite{candes2006stable} that describes the $\ell_2$ error rate of the BP algorithm. Informally, the result assumes that the noise is bounded as $\|\e\|_2\leq \e_0$ for some constant $\e_0>0$,  the matrix $A\in \R^{n\times p}$ is drawn from a suitable ensemble, and the number of measurements satisfies $n \gtrsim T\log (p/T)$ for some $T\in \{1,\dots, p\}$. The conclusion is that with high probability, the solution \mbox{$\hat{x}\in \text{argmin} \{ \|v \|_1 : \|Av - y\|_2\leq \e_0, v\in \R^p\}$} satisfies
\begin{equation}\label{candesTao}
\|\hat{x}-x\|_2 \leq c_{1}\,\e_0 + c_{2} \,\ts\frac{\|x- x_{T}\|_1}{\sqrt{T}},
\end{equation}
where $x_{T}\in \R^p$ is the best $T$-term approximation\footnote{The vector $x_T\in\R^p$ is obtained by setting to 0 all coordinates of $x$ except the $T$ largest ones (in magnitude). } to $x$, and $c_1,c_2>0$ are constants. This bound is a fundamental point of reference, since it matches the minimax optimal rate under certain conditions, and applies to all signals $x\in \R^p$ (rather than just $k$-sparse signals). Additional details may be found in~\cite{caiRIP}\,[Theorem 3.3],~\cite{vershyninIntro}\,[Theorem 5.65].

We  now aim to answer the question, ``If $s(x)$ were known, how large must $n$ be in order for $\|\hat{x}-x\|_2$ to be small?''. To make the choice of $n$ independent of the scale of $x$, we consider the \emph{relative} $\ell_2$ error 
\begin{equation}\label{scaled}
\ts\frac{\|\hat{x}-x\|_2}{\|x\|_2} \leq c_{1}\,\ts\frac{\e_0}{\|x\|_2} + c_{2} \,\ts\frac{1}{\sqrt{T}}\frac{\|x- x_{T}\|_1}{\|x\|_2},
\end{equation}
so that the \emph{approximation error} term $\frac{1}{\sqrt{T}}\frac{\|x- x_{T}\|_1}{\|x\|_2}$ is invariant to the transformation $x\mapsto c x$ with $c\neq 0$. Also, since the noise-to-signal ratio $\e_0/\|x\|_2$ is a fixed feature of the problem, the choice of $n$ is determined only by the approximation error.
Since the bound~\eqref{candesTao} assumes $n\gtrsim T \log(p/T)$, our question is reduced to determining how large $T$ must be relative to $s(x)$ in order for $\frac{1}{\sqrt{T}}\frac{\|x- x_{T}\|_1}{\|x\|_2}$ to be small.
Proposition~\ref{recovCond} shows that the condition $T \gtrsim  s(x)$
is necessary for the approximation error to be small, and the condition $T\gtrsim s(x)\log(p)$ is sufficient.

~\\
%
%
%
%
%
\begin{proposition}\label{recovCond}
 Let $x\in\R^p\setminus\{0\}$, and $T\in\{1,\dots,p\}$. The following statements hold for any $c,\varepsilon>0.$
 \begin{enumerate}[(i)]
 \vspace{-.2cm}
\item \label{slower} If the $T$-term approximation error satisfies $\frac{1}{\sqrt{T}} \frac{\|x-x_T\|_1}{\|x\|_2} \leq \ve$, then
$$ T \geq \ts\frac{1}{(1+\ve)^2}\cdot s(x).$$
\vspace{-.3cm}
\item\label{supper} If  $T\geq  c\,s(x)\log(p)$, then the $T$-term approximation error satisfies
$$ \ts\frac{1}{\sqrt{T}} \frac{\|x-x_T\|_1}{\|x\|_2} \leq \ts\frac{1}{\sqrt{c_0\log(p)}} \big(1-\ts\frac{T}{p}\big).$$
In particular,  if $T\geq 2 s(x) \log(p)$ with $p\geq 100$, then 
$$\ts\frac{1}{\sqrt{T}} \frac{\|x-x_T\|_1}{\|x\|_2} \leq \ts \ts\frac{1}{3}.$$
\end{enumerate}
\end{proposition}

\paragraph{Remarks.} A notable feature of these bounds is that they hold for all non-zero signals.
 In our simulations in Section~\ref{sec:sims}, we show that choosing  $n= 2 \lceil \hat{s}(x)\rceil \log(p/\lceil \hat{s}(x)\rceil)$ based on an estimate $\hat{s}(x)$ leads to accurate reconstruction across many sparsity levels.
\section{Estimation results for $s(x)$}\label{sec:est}

 In this section, we present a simple procedure to estimate $s(x)$ for any $x\in\R^p\setminus\{0\}$. The procedure uses a small number of measurements, makes no sparsity assumptions, and requires very little computation. The measurements we prescribe may also be re-used to recover the full signal after the parameter $s(x)$ has been estimated.

The results in this section are are based on the measurement model~\eqref{setup1}, which may be written in scalar notation as
 \begin{equation}\label{model}
y_i = \langle a_i,x\rangle +\e_i, \ \ \ \ i=1,\dots, n.
\end{equation} 
We assume the noise variables $\e_1,\dots,\e_n$ are independent, and bounded by $|\e_i|\leq \sigma_0$, for some constant $\sigma_0>0$. No additional structure on the noise is needed.
\subsection{Sketching with stable laws}
Our estimation procedure derives from a technique known as \emph{sketching} in the streaming computation literature~\cite{indykstable}. Although this area deals with problems that are mathematically similar to CS, the connections between the two areas do not seem to be fully developed. One reason for this may be that measurement noise does not usually play a role in the context of sketching.

For any $q\in (0,2]$, the sketching technique offers a way to estimate $\|x\|_q$ from a set of randomized linear measurements.
  In our approach, we estimate $s(x)=\|x\|_1^2/\|x\|_2^2$ by estimating $\|x\|_1$ and $\|x\|_2$ from separate sets of measurements.
The core idea is to generate the measurement vectors $a_i\in\R^p$ using \emph{stable laws}~\cite{zolotarev,indykstable}.
\begin{definition}\label{def:stable}
A random variable $V$ has a \emph{symmetric stable distribution} if its characteristic function is of the form $\E[\exp(\sqrt{-1}tV)] = \exp(-|\gamma t|^{q})$ for some $q\in(0,2]$ and some $\gamma>0$.  We denote the distribution by $V\sim S_{q}(\gamma)$, and $\gamma$ is referred to as the \emph{scale} parameter.
 \end{definition}
 The most well-known examples of symmetric stable laws are the cases
 of $q=2,1$, namely the Gaussian distribution $N(0,\gamma)=S_2(\gamma)$, and  the Cauchy distribution $C(0,\gamma)=S_1(\gamma)$. To fix some notation, if a vector $a_1=(a_{1,1},\dots,a_{1,p})\in\R^p$ has i.i.d.\!\! entries drawn from $S_{q}(\gamma)$, we write $a_1\sim S_{q}(\gamma)^{\otimes p}$. The connection with $\ell_q$ norms hinges on the following property of stable distributions~\cite{zolotarev}.
\begin{fact} Suppose $x\in\R^p$, and $a_1\sim S_q(\gamma)^{\otimes p}$ with parameters $q\in (0,2]$ and $\gamma>0$. Then, the random variable $\langle x,a_1\rangle$ is distributed according to $S_q(\gamma^q\|x\|_{q}^{q}).$
\end{fact}
Using this fact, if we generate a set of i.i.d.\!\! vectors $a_1,\dots,a_n$ from $S_q(\gamma)^{\otimes p}$ and let $y_i=\langle a_i,x\rangle$, then $y_1,\dots,y_n$ is an i.i.d.\!\! sample from $S_q(\gamma^q\|x\|_{q}^{q})$. Hence, in the special case of noiseless linear measurements, the task of estimating $\|x\|_q$ is equivalent to a well-studied univariate problem: \emph{estimating the scale parameter of a stable law from an i.i.d. sample}.\footnote{It is worth pointing out that the $\ell_0$ norm $\|x\|_0$ can be estimated as the limit of $\|x\|_q^q$ with $q\downarrow 0$. This can be useful in the streaming computation context where $x$ often represents a sparse stream of \emph{integers}~\cite{cormode}. However, we do not pursue this approach, since $\|x\|_0$ less meaningful for natural signals. }

When the $y_i$ are corrupted with noise, our analysis shows that standard estimators for scale parameters are only moderately affected. The impact of the noise can also be reduced via the choice of $\gamma$ when generating $a_i\sim S_q(\gamma)^{\otimes p}$. The $\gamma$ parameter controls the ``energy level'' of the measurement vectors $a_i$.  (Note that in the Gaussian case, if $a_1\sim S_2(\gamma)^{\otimes p}$, then $\E\|a_1\|_2^2  = \gamma^2 p$.) In our results, we leave $\gamma$ as a free parameter to show how the effect of noise is reduced as $\gamma$ is increased.

 \subsection{Estimation procedure for $s(x)$}\label{sec:procedure}

Two sets of measurements are used to estimate $s(x)$, and we write the total number as $n=n_1+n_2$. The first set is obtained by generating i.i.d.\!\! measurement vectors from a Cauchy distribution, 
\begin{equation}\label{cauchymeas}
a_i  \sim C(0,\gamma)^{\otimes p}, \ \ \ \ i=1,\dots, n_1. 
\end{equation}
The corresponding values $y_i$ are then used to estimate $\|x\|_1$ via the statistic
\begin{equation}\label{cauchystat}
\hat{T}_1:=\ts\frac{1}{\gamma}\text{median}(|y_1|,\dots,|y_{n_1}|),
\end{equation}
which is a standard estimator of the scale parameter of the Cauchy distribution~\cite{fama,pingliJMLR}. Next, a second set of i.i.d.\!\! measurement vectors are generated from a Gaussian distribution
\begin{equation}\label{gaussmeas}
\ \ \ \ \ \ \ \ a_i  \sim N(0,\gamma)^{\otimes p}, \ \ \ \ i=n_1+1,\dots, n_1+n_2. 
\end{equation}
In this case, the associated $y_i$ values are used to compute an estimate of $\|x\|_2^2$ given by 
\begin{equation}\label{gaussstat}
\hat{T}_2^2:={\ts\frac{1}{\gamma^2 n_2 }}(y_{n_1+1}^2+\cdots+y_{n_1+n_2}^2),
\end{equation}
 which is the natural estimator of the scale parameter (variance) of a Gaussian distribution. Combining these two statistics, our estimate of $s(x)=\|x\|_1^2/\|x\|_2^2$ is defined as
\begin{equation}\label{shat}
\hat{s}(x):=\hat{T}_1^2\big/\hat{T}_2^2.
\end{equation}
\subsection{Confidence interval.}\label{sec:CI}
 The following theorem describes the relative error $\big|\frac{\hat{s}(x)}{s(x)}-1\big|$ via an asymptotic confidence interval for $s(x)$. Our result is stated in terms of the noise-to-signal ratio 
 $$\rho:=\ts\frac{\sigma_0}{\gamma \|x\|_2},$$
 and the standard Gaussian quantile  $z_{1-\alpha}$, which satisfies $\Phi(z_{1-\alpha})=1-\alpha$ for any coverage level $\alpha\in(0,1)$. In this notation, the following parameters govern the width of the confidence interval,
 \begin{align*}
 \ts
 \eta_n(\alpha,\rho) :=\ts\frac{ z_{1-\alpha}}{\sqrt{n}}+\rho \text{ \ \ and \ \ }\delta_n(\alpha,\rho) :=\ts \frac{ \pi z_{1-\alpha}}{\sqrt{2n}}+\rho,
 \end{align*}
 and we write these simply as $\delta_n$ and $\eta_n$.
As is standard in high-dimensional statistics, we allow all of the model parameters $p, x, \sigma_0$ and $\gamma$ to vary implicitly as functions of $(n_1,n_2)$, and let $(n_1,n_2)\to\infty$.   For simplicity, we choose to take measurement sets of equal sizes, $n_1=n_2=n/2$, and we place a mild constraint on $\rho$, namely $\eta_n(\alpha,\rho)<1$. (Note that standard algorithms such as Basis Pursuit are not expected to perform well unless $\rho\ll 1$, as is clear from the bound~\eqref{scaled}.) Lastly, we make no restriction on the growth of $p/n$, which makes $\hat{s}(x)$ well-suited to high-dimensional problems.
%
%
\begin{theorem}\label{ThmNoise} Let $\alpha\in (0,1/2)$ and $x\in\R^p\setminus\{0\}$. Assume that $\hat{s}(x)$ is constructed as above, and that the 
model~\eqref{model} holds.  Suppose also that $n_1=n_2=n/2$ and $\eta_n(\alpha,\rho)<1$ for all $n$.
Then as $n\to\infty$, we have
\begin{equation}\label{thmmain}
\P\Big(\ts\sqrt{\frac{\hat{s}(x)}{s(x)}} \in \big[\frac{1-\delta_n}{1+\eta_n},\frac{1+\delta_n}{1-\eta_n}\big]\Big) \geq (1-2\alpha)^2+o(1).
\end{equation}
\end{theorem}
%
\paragraph{Remarks.} The most important feature of this result is 
that the width of the confidence interval does \emph{not} depend on the dimension or sparsity of the 
unknown signal. Concretely, this means that the number of measurements needed to estimate $s(x)$ to a fixed precision is only $\mathcal{O}(1)$ with respect to the size of the problem. This conclusion is 
confirmed our simulations in Section~\ref{sec:sims}. Lastly, when $\delta_n$ and $\eta_n$ are small, we note that the relative error $|\hat{s}(x)/s(x)-1|$ is of order $(n^{-1/2}+\rho)$ with high probability, which follows from the simple Taylor expansion $\frac{(1+\ve)^2}{(1-\ve)^2} = 1+4\ve+o(\ve)$.

\subsection{Estimating rank and sparsity of matrices}\label{matrixExt}
The framework of CS naturally extends to the problem of recovering 
an unknown matrix \mbox{$X\in \R^{p_1\times p_2}$} on the basis of the 
measurement model
\begin{equation}\label{matrixModel}
\ts y = \A(X) + \e,
\end{equation}
where $y\in \R^n$,
$\A$ is a user-specified linear operator from $\R^{p_1\times p_2}$ to $\R^n$, and 
 $\e\in\R^n$ is a vector of noise variables. 
In recent years, many researchers have explored the recovery of $X$
when it is assumed to have sparse or low rank structure. We refer to the papers~\cite{candesplan,venkatinverse}
for descriptions of numerous applications.
In analogy with the previous section, the parameters $\text{rank}(X)$ or $\|X\|_0$ play important theoretical roles, but are very sensitive to perturbations of $X$. Likewise, it is of basic interest to estimate robust measures of rank and sparsity for matrices. Since the sparsity analogue $s(X):=\|X\|_1^2/\|X\|_F^2$ can be estimated as a straightforward extension of Section~\ref{sec:procedure}, we restrict our attention to the more distinct problem of rank estimation.

 %
 %
 \subsubsection{The rank of semidefinite matrices}\label{sec:psd}
 
In the context of recovering a low-rank positive semidefinite matrix 
$X\in \mathbb{S}_+^{p\times p}\backslash\{0\}$, 
the quantity $\text{rank}(X)$ plays the role that the 
norm $\|x\|_0$ does in the recovery of a sparse vector. If we let 
$\lambda(X)\in\R^p_+$ denote the vector of ordered eigenvalues of 
$X$, the connection can be made explicit by writing
$\text{rank}(X)=\|\lambda(X)\|_0$. As in Section~\ref{sec:procedure}, 
our approach is to consider a robust alternative to the rank. Motivated 
by the quantity $s(x)=\|x\|_1^2\big/\|x\|_2^2$ in the vector case, 
we now consider 
$$\ts r(X):=\frac{\|\lambda(X)\|_1^2}{\|\lambda(X)\|_2^2} = \frac{\tr(X)^2}{\|X\|_F^2}$$
as our measure of the effective rank for non-zero $X$, which always satisfies $1\leq r(X)\leq p$. The quantity $r(X)$
has appeared elsewhere as a measure of rank~\cite{lopes,ellStar}, but is less well known than other rank relaxations, such as the \emph{numerical rank}
 $\|X\|_F^2\big/\|X\|\op^2$~\cite{vershyninnumerical}. 
The relationship between $r(X)$ and $\text{rank}(X)$ is completely 
analogous to $s(x)$ and $\|x\|_0$. Namely, 
we have a sharp, scale-invariant inequality
$$r(X)\leq \text{rank}(X),$$
with equality holding iff the non-zero eigenvalues of $X$ are equal. 
The quantity $r(X)$ 
is more stable than $\text{rank}(X)$ in the sense that if $X$ has $k$ large
eigenvalues, and $p-k$ small eigenvalues, then 
$r(X)\approx k$, whereas $\text{rank}(X)=p$. 

Our procedure for estimating $r(X)$ is based on estimating $\tr(X)$ 
and $\|X\|_F^2$ from separate sets of measurements. The semidefinite condition is exploited through the basic relation
\mbox{$\langle I_{p\times p},X\rangle =\tr(X)=\|\lambda(X)\|_1$}.
 To estimate $\tr(X)$, we 
use $n_1$ linear measurements of the form
\begin{equation}\label{tracemeas1}
y_i = \langle \gamma I_{p\times p},X\rangle +\e_i,  \ \ \ \ i=1,\dots, n_1
\end{equation}
and compute the estimator
$\ts \breve{T}_1:=\ts\frac{1}{\gamma}\frac{1}{n_1}\sum_{i=1}^{n_1} y_i$, where $\gamma>0$ is again the measurement energy parameter. Next, to estimate 
$\|X\|_F^2$, we note that if $Z\in \R^{p\times p}$ has i.i.d. $N(0,1)$ entries, then $\E\langle X,Z\rangle^2=\|X\|_F^2$. Hence, if we collect $n_2$ additional measurements of the form
\begin{equation}\label{tracemeas2}
y_i= \langle \gamma Z_i,X \rangle +\e_i, \ \ \ \ i=n_1+1,\dots,n_1+n_2,
\end{equation}
where the $Z_i\in \R^{p\times p}$ are independent random matrices 
with i.i.d. $N(0,1)$ entries, then a suitable estimator of $\|X\|_F^2$ is
$\ts \breve{T}_2^2:=\frac{1}{\gamma^2 n_2}\sum_{i=n_1+1}^{n_1+n_2} y_i^2$. 
Combining these 
statistics, we propose
$$\ts \hat{r}(X):=\breve{T}_1^2\big/\breve{T}_2^2$$
as our estimate of $r(X)$. In principle, this procedure can be refined by using the measurements~\eqref{tracemeas1} to estimate the noise distribution, but we omit these details. Also for simplicity, we retain the assumptions of the previous section, and assume only that the $\e_i$ are independent, and bounded by $|\e_i|\leq \sigma_0$.  The next theorem shows that the estimator $\hat{r}(X)$
mirrors $\hat{s}(X)$ as in Theorem~\ref{ThmNoise}, but with $\rho$ being replaced by $\varrho:=\sigma_0\big/(\gamma \|X\|_F)$,  and with $\eta_n$ being replaced by $\zeta_n=\zeta_n(\varrho,\alpha):= z_{1-\alpha}/\sqrt{n}+\varrho$.

\begin{theorem}\label{RankNoise} Let $\alpha\in (0,1/2)$ and $X\in\mathbb{S}_+^{p\times p}\setminus\{0\}$. Assume that $\hat{r}(X)$ is constructed as above, and that the 
model~\eqref{matrixModel} holds.  Suppose also that $n_1=n_2=n/2$ and $\zeta_n(\alpha,\rho)<1$ for all $n$.
Then as $n\to\infty$, we have
\begin{equation}\label{thmmain}
\P\Big(\ts\sqrt{\frac{\hat{r}(X)}{r(X)}} \in \big[\frac{1-\varrho}{1+\zeta_n},\frac{1+\varrho}{1-\zeta_n}\big]\Big) \geq 1-2\alpha+o(1).
\end{equation}
\end{theorem}

\paragraph{Remarks.} In parallel with Theorem~\ref{ThmNoise}, this confidence interval has the valuable property that its width does not depend on the rank or dimension of $X$, but merely on the noise-to-signal ratio $\varrho=\sigma_0\big/(\gamma \|X\|_F)$. The relative error $|\hat{r}(X)/r(X)-1|$ is again of order $(n^{-1/2}+\varrho)$ with high probability when $\zeta_n$ is small.
%
%
%
%
%
 %
 %
 %

\section{Deterministic measurement matrices}\label{sec:neg}
The problem of constructing deterministic matrices $A$ with good recovery properties (e.g. RIP-$k$ or NSP-$k$) has been a longstanding topic within CS. Since our procedure in Section~\ref{sec:procedure} selects $A$ at random, it is natural to ask if randomization is essential to the estimation of unknown sparsity. In this section, we show that estimating $s(x)$ with a deterministic matrix $A$ leads to results that are inherently different from our randomized procedure.

At an informal level, the difference between random and deterministic matrices makes sense if we think of the estimation problem as a game between nature and a statistician. Namely, the statistician first chooses a matrix $A\in\R^{n\times p}$ and an estimation rule $\delta :\R^n\to \R$. (The function $\delta$ takes $y\in\R^n$ as input and returns an estimate of $s(x)$.)  In turn, nature chooses a signal $x\in\R^p\setminus\{0\}$, with the goal of maximizing the statistician's error. When the statistician chooses $A$ deterministically, nature has the freedom to adversarially select an $x$ that is ill-suited to the fixed matrix $A$. By contrast, if the statistician draws $A$ at random, then nature does not know what value $A$ will take, and therefore has less knowledge to choose a bad signal.

In the case of noiseless \emph{random} measurements, Theorem~\ref{ThmNoise} implies that our particular estimation rule $\hat{s}(x)$ can achieve a relative error on the order of \mbox{$|\hat{s}(x)/s(x) -1|=\mathcal{O}(n^{-1/2})$} with high probability for any non-zero $x$.  (cf. Remarks for Theorem~\ref{ThmNoise}.) Our aim is now to show that for noiseless \emph{deterministic} measurements, \emph{all} estimation rules $\delta$ have a worst-case relative error $|\delta(Ax)/s(x)-1|$ that is much larger than than $n^{-1/2}$. In other words, there is always a choice of $x$ that can defeat a deterministic procedure, whereas $\hat{s}(x)$ is likely to succeed under any choice of $x$. 

In stating the following result, we note that it involves no randomness whatsoever --- since we assume that the observed measurements $y=Ax$ are noiseless and obtained from a deterministic matrix $A$.
\vspace{0.06cm}
 \begin{theorem}\label{minimaxnew}
 The minimax relative error for estimating $s(x)$ from noiseless deterministic measurements $y=Ax$ satisfies
\begin{equation*}\label{minmaxnew}
 \inf_{A\in \R^{n\times p}} \inf_{\delta:\R^n\to\R} \:\sup_{x\in\R^p\setminus\{0\}}\Big|\ts\frac{\delta(Ax)}{s(x)}-1\Big|\geq \ts \frac{1-(n+1)/p}{2(1+2\sqrt{2 \log(2p)})^2}.
 \end{equation*}
\end{theorem}
\paragraph{Remarks.} Under the typical high-dimensional scenario where there is some $\kappa\in (0,1)$ for which $n/p\to \kappa$ as $(n,p)\to\infty$,  we see that $ |\ts\frac{\delta(Ax)}{s(x)}-1| \gtrsim \ts\frac{1}{\log(n)},$
 which is indeed much larger than $n^{-1/2}$.

\section{Simulations}\label{sec:sims}
\begin{figure*}[b!]
\centering
{\includegraphics[angle=0,
  width=.31\linewidth]{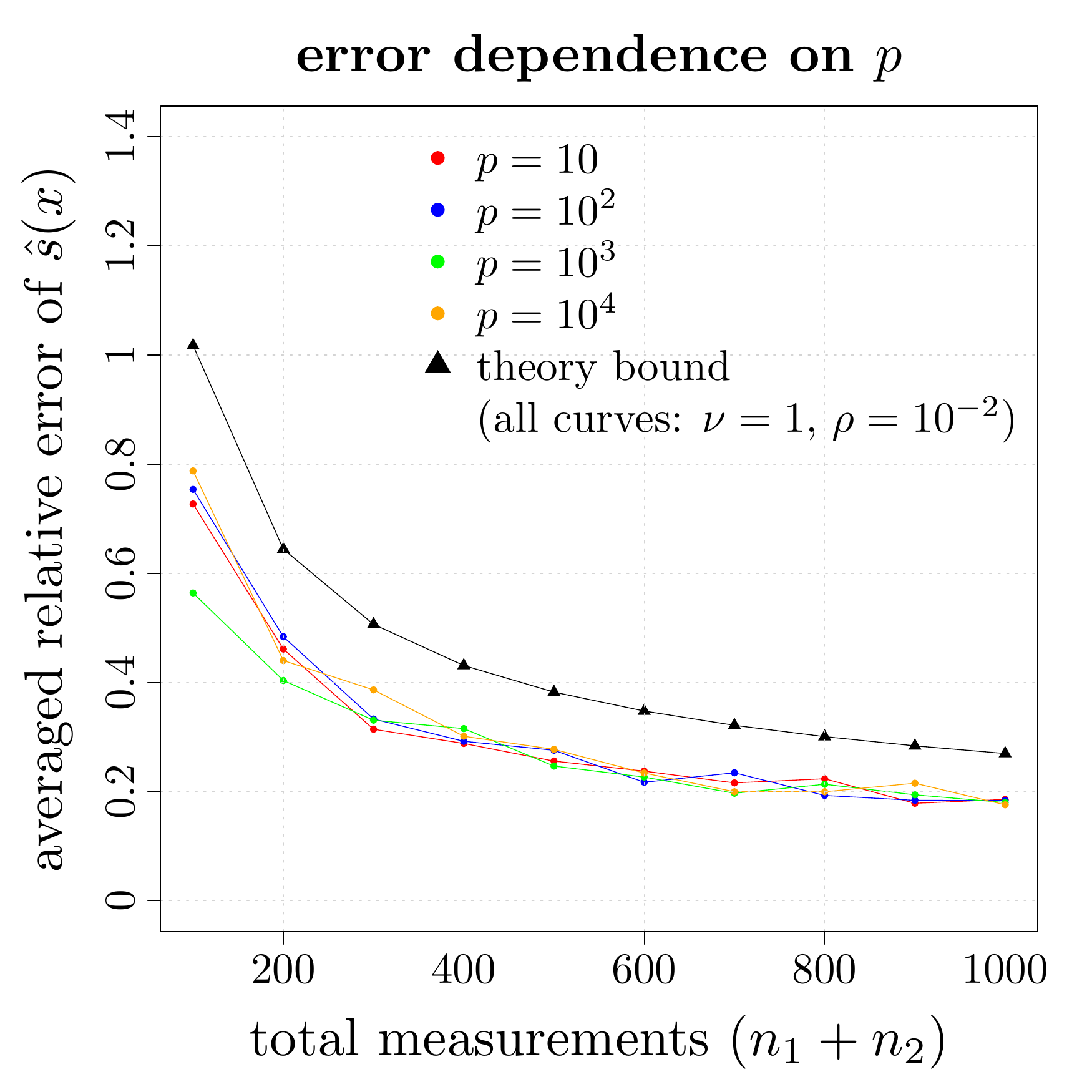}}
{\includegraphics[angle=0,
  width=.31\linewidth]{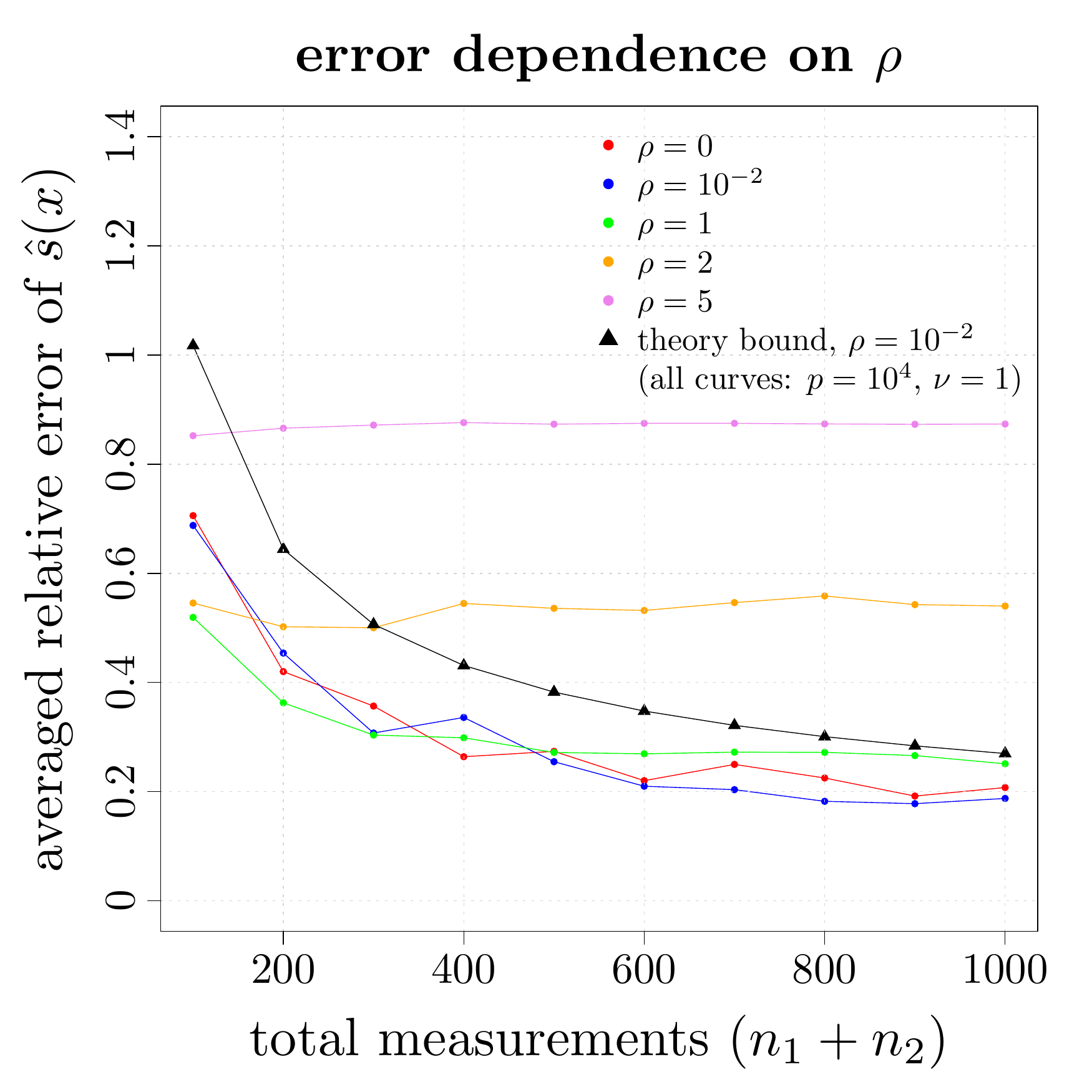}}
{\includegraphics[angle=0,
  width=.31\linewidth]{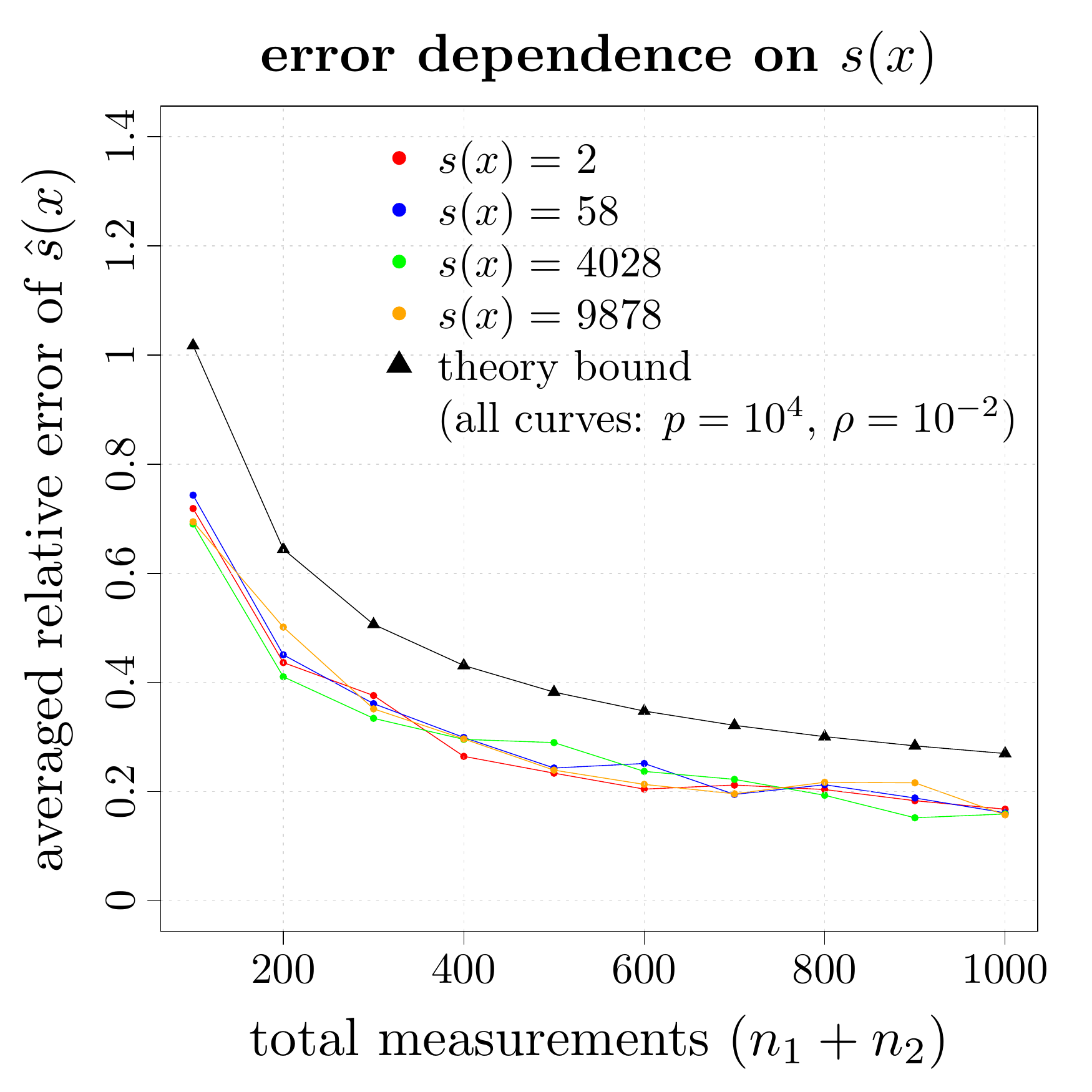}}
\caption{Performance of $\hat{s}(x)$ as a function of $p$, $\rho$, $s(x)$, and number of measurements.
}
\label{fig:shatPlot}
\end{figure*}

\paragraph{Relative error of $\hat{s}(x)$.} To validate the consequences of Theorem~\ref{ThmNoise}, we study how the decay of the relative error $|\hat{s}(x)/s(x)-1|$ depends on the parameters $p$, $\rho$, and $s(x)$. We generated measurements $y=Ax+\e$ under a broad range of parameter settings, with $x\in\R^{10^4}$  in most cases. Note that although $p=10^4$ is a very large dimension, it is not at all extreme from the viewpoint of applications (e.g. a megapixel image with $p=10^6$). Additional details regarding parameters are given below.
As anticipated by Theorem~\ref{ThmNoise}, the left and right panels in Figure~\ref{fig:shatPlot} show that the relative error has no noticeable dependence on $p$ or $s(x)$. The middle panel shows that for fixed $n_1+n_2$, the relative error grows moderately with $\rho=\ts\frac{\sigma_0}{\gamma \|x\|_2}$. Lastly, our theoretical bounds on $|\hat{s}(x)/s(x)-1|$ conform well with the empirical curves in the case of low noise ($\rho=10^{-2}$).  
\paragraph{Reconstruction of $x$ based on $\hat{s}(x)$.} For the problem of choosing the number of measurements $n$ for reconstructing $x$,  we considered the choice of $\hat{n}:=2\lceil \hat{s}(x)\rceil \log(p/\lceil \hat{s}(x)\rceil)$. First, to compute $\hat{s}(x)$, we followed Section~\ref{sec:procedure}, and drew initial measurement sets of Cauchy and Gaussian vectors with $n_1=n_2=500$ and $\gamma=1$. If it happened to be the case that  $500\geq \hat{n}$, then reconstruction was performed using only the initial 500 Gaussian measurements. Alternatively, if $\hat{n}> 500$, then $(\hat{n}-500)$ additional Gaussian measurements were drawn for reconstruction. 
 Specific details of the implementation with the Basis Pursuit algorithm are given below.  Figure 3 illustrates the results for three power-law signals in $\R^{10^4}$ with $x_{[i]}\propto i^{-\nu}$,  $\nu=0.7, 1.0, 1.3$ and $\|x\|_2=1$ (corresponding to $s(x)=823, 58, 11$). In each panel, the coordinates of $x$ are plotted in black, and those of  $\hat{x}$ are plotted in red. Clearly, there is good qualitative agreement in all three cases.
 From left to right, the value of $\hat{n}=2\lceil \hat{s}(x)\rceil \log(p/\lceil \hat{s}(x)\rceil)$ was 4108, 590, and 150. Altogether, the simulation indicates that $\hat{n}$ takes the structure of the true signal into account, and is sufficiently large for accurate reconstruction.
\begin{figure*}[b!]
\centering
{\includegraphics[angle=0,
  width=.31\linewidth]{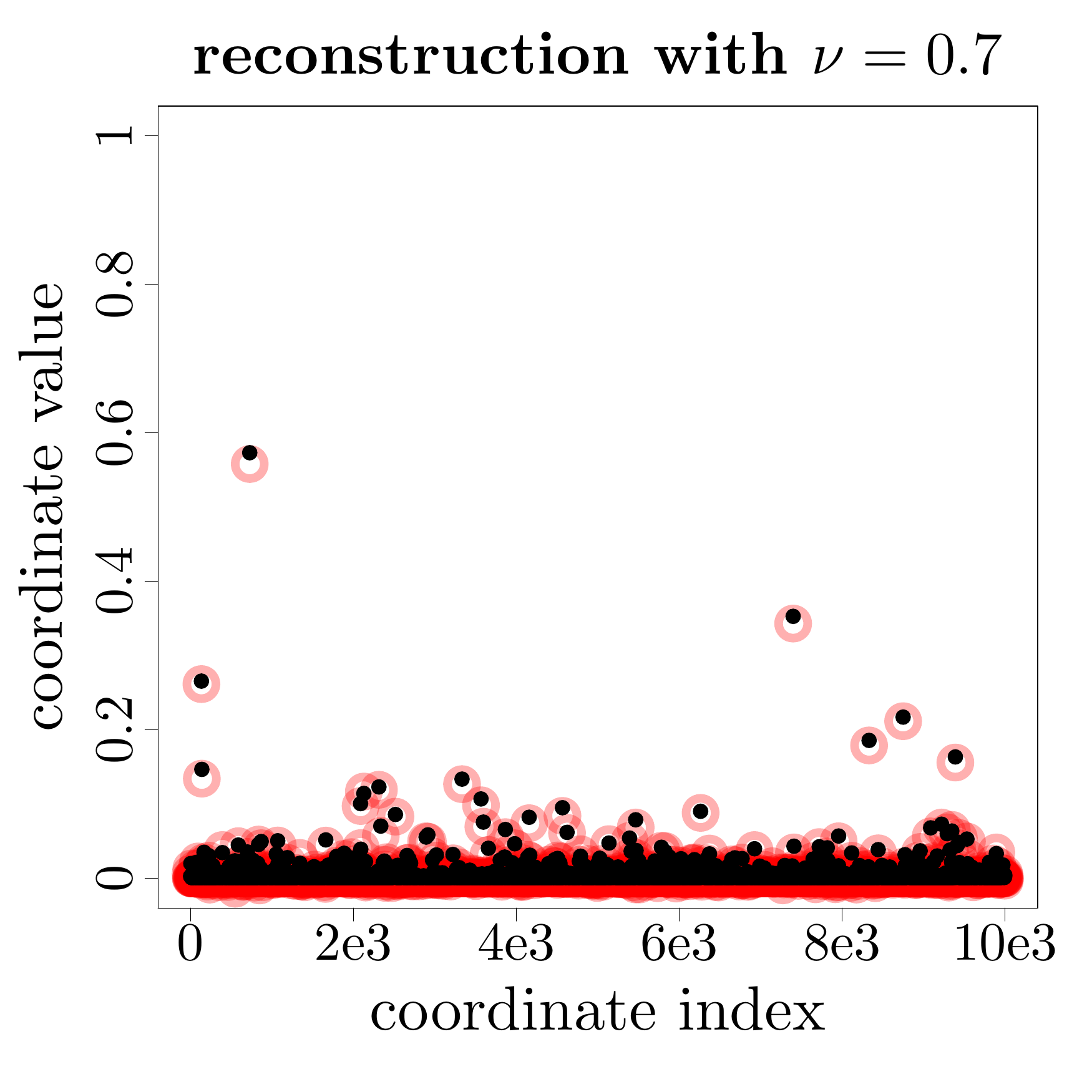}}
{\includegraphics[angle=0,
  width=.31\linewidth]{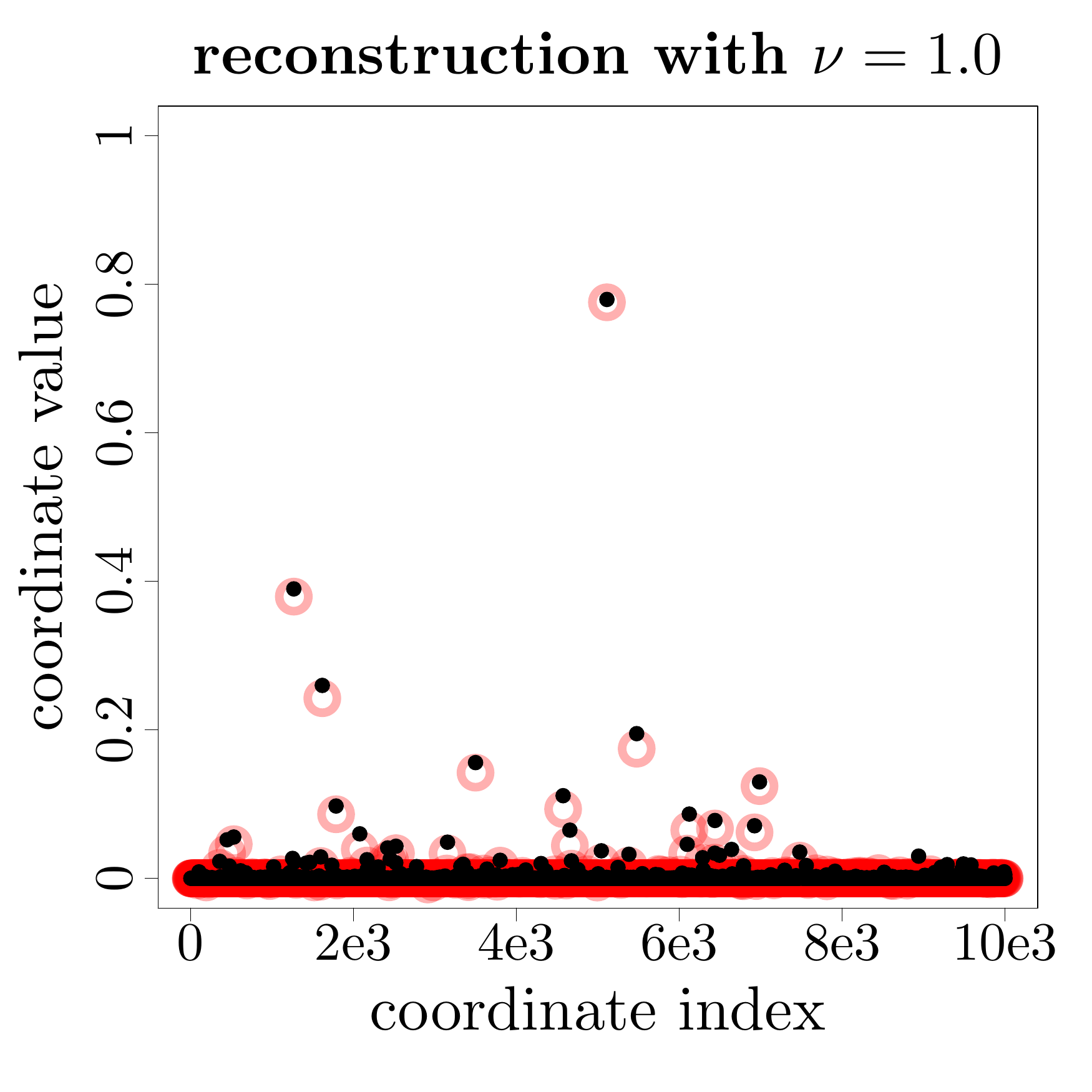}}
{\includegraphics[angle=0,
 width=.31\linewidth]{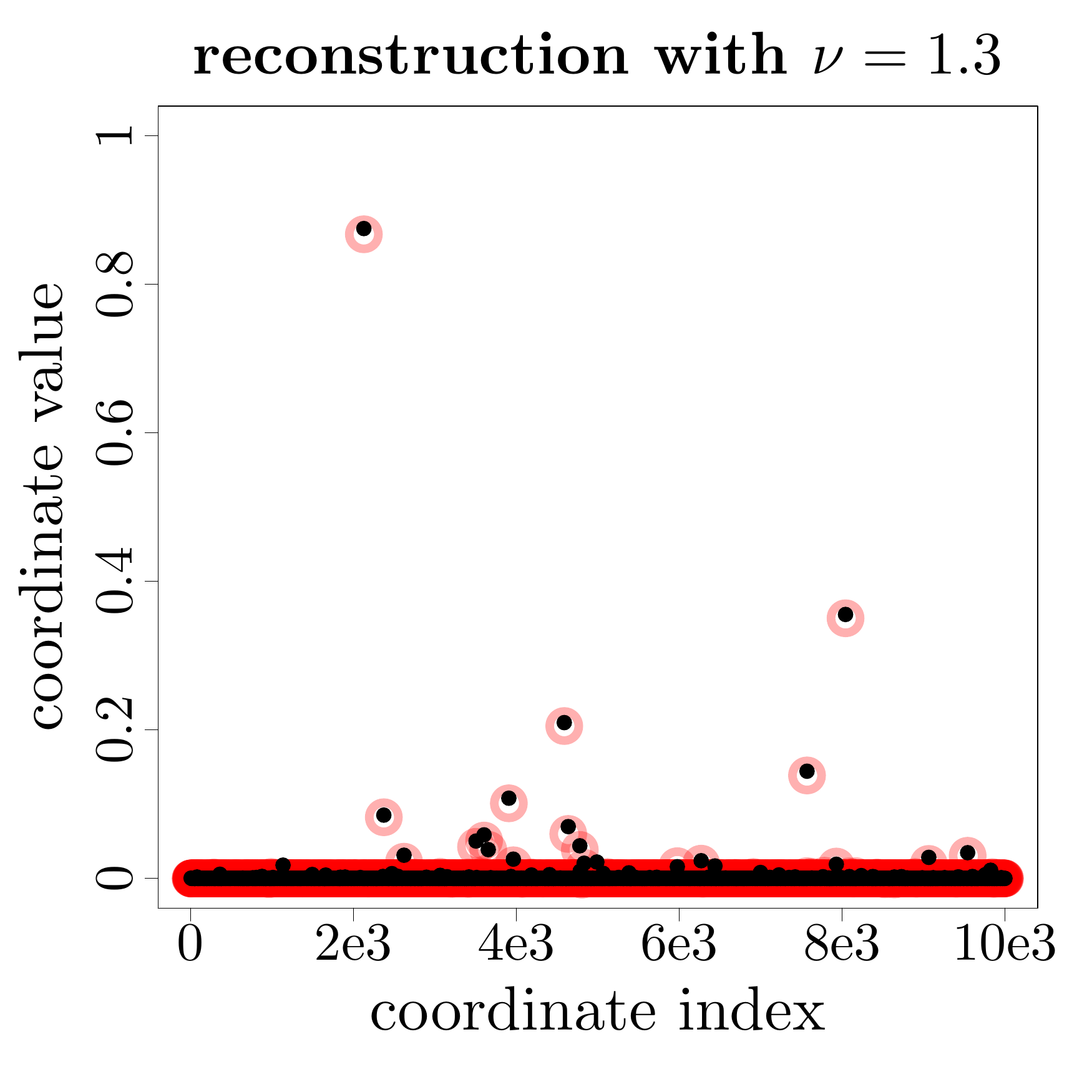}}
\caption{Signal recovery after choosing $n$ based on $\hat{s}(x)$. True signal $x$ in black, and $\hat{x}$ in red.}

\label{fig:recovPlot}
\end{figure*}

\paragraph{Settings for relative error of $\hat{s}(x)$ (Figure 2).}
For each parameter setting labeled in the figures, we let $n_1=n_2$ and averaged \mbox{$|\hat{s}(x)/s(x)-1|$} over 100 problem instances of $y=Ax+\e$. In all cases, the matrix $A$ chosen according to equations~\eqref{cauchymeas} and~\eqref{gaussmeas} with $\gamma=1$ and $\e \sim \text{Uniform}[-\sigma_0,\sigma_0]$. We always chose the normalization $\|x\|_2=1$, and $\gamma=1$ so that $\rho=\sigma_0$. (In the left and right panels, $\rho =10^{-2}$.)  For the left and middle panels, all signals have the decay profile $x_{[i]}\propto i^{-1}$. For the right panel, the values $s(x)=$2, 58, 4028, and 9878 were obtained using decay profiles $x_i \propto i^{-\nu}$ with $\nu=2,1,1/2,1/10$.  In the left and right panels, we chose $p=10^4$ for all curves. The theoretical bound on $|\hat{s}(x)/s(x)-1|$ in black was computed from Theorem~\ref{thmmain} with $\alpha=0.25$. (This bound holds with probability at least 1/2 equal to $1/2$ and so it may be reasonably be compared with the average of $|\hat{s}(x)/s(x)-1|$.)

\paragraph{Settings for reconstruction (Figure 3).}
Reconstructions were computed using the BP algorithm \mbox{$\hat{x}\in \text{argmin}\{\|v\|_1:\|Av-y\|_2\leq \e_0, v\in\R^p \}$}, with the choice $\e_0=\sigma_0\sqrt{\hat{n}}$ being based on i.i.d. noise variables $\e_i\sim \text{Uniform}[-\sigma_0,\sigma_0]$, with $\sigma_0=0.001$. (As defined above, $\hat{n}$ is the instance-dependent number of Gaussian measurements.) For each choice of $\nu=0.7,1.0,1.3$, we generated 25 problem instances and plotted the reconstruction corresponding to the median of $\|\hat{x}-x\|_2$ over the 25 runs (so that the plots reflect typical performance). 

\section{Appendix}

This appendix contains the proofs of Proposition~\ref{recovCond}, as well as Theorems~\ref{ThmNoise},~\ref{RankNoise}, and~\ref{minimaxnew}.

\subsection{Proof of Proposition 1}

To prove the implication \emph{(i)},  we calculate
\begin{equation}
\begin{split}\label{lowerbound}
\ts \frac{1}{\sqrt{T}} \frac{\|x-x_T\|_1}{\|x\|_2}&=\ts \frac{1}{\sqrt{T}} \frac{\|x\|_1-\|x_T\|_1}{\|x\|_2}\\
&=\ts \frac{\sqrt{s(x)}}{\sqrt{T}} - \frac{1}{\sqrt{T}} \frac{\|x_T\|_1}{\|x_T\|_2}\frac{\|x_T\|_2}{\|x\|_2}\\
&=\ts \frac{\sqrt{s(x)}}{\sqrt{T}} -\frac{\sqrt{s(x_T)}}{\sqrt{T}}\frac{\|x_T\|_2}{\|x\|_2}.
\end{split}
\end{equation}
Since $s(x_T)\leq \|x_T\|_0\leq T$, and $\frac{\|x_T\|_2}{\|x\|_2}\leq 1$, we obtain the lower bound
$$\ts \frac{1}{\sqrt{T}} \frac{\|x-x_T\|_1}{\|x\|_2} \geq \ts \frac{\sqrt{s(x)}}{\sqrt{T}} - 1.$$
Hence, if the left hand side is at most $\ve$, we must have $T\geq\ts \frac{s(x)}{(1+\ve)^2}$, proving \emph{(i)}.

\noindent To prove the second implication, note that $T \geq \ts c \log(p) \frac{\|x\|_1^2}{\|x\|_2^2}$ implies
\begin{equation}\label{basic}
\begin{split}
\ts \frac{1}{\sqrt{T}}\frac{\|x-x_T\|_1}{\|x\|_2}&\leq  \ts\frac{1}{\sqrt{c\log(p)}}\frac{\|x-x_T\|_1}{\|x\|_1}\\
& =  \ts\frac{1}{\sqrt{c\log(p)}}\left(1- \ts\frac{\|x_T\|_1}{\|x\|_1}\right).
\end{split}
\end{equation}
 Next, consider the probability vectors $u,v\in\R^p$ defined by $u_i = 1/p$ and $v_i = |x|_{[i]}/\|x\|_1$ (that is, $v_1\geq v_2\geq\cdots \geq v_p$). It is a basic fact about the majorization ordering on $\R^p$ that $u$ is majorized by every probability vector~\cite[p. 7]{olkin}. In particular, we have $\sum_{i=1}^T u_i \leq \sum_{i=1}^T v_i$ for any $T\in \{1,\dots,p\}$, which is the same as
$$ \ts\frac{T}{p} \leq \ts\frac{\|x_T\|_1}{\|x\|_1}.$$
Combining this with line~\eqref{basic} proves \emph{(ii)}. \qed 
\subsection{Proof of Theorem 1.}

Define the noiseless version of the measurement $y_i$ to be
$$ \ \ \ \ \ \ \ \ \ \ \  \ \ \ \ \ \ \ \ \ \ \ \ y_i\ccirc := \langle a_i, x\rangle,  \ \ \ \ \ i=1,\dots,n_1+n_2 $$
and let the noiseless versions of the statistics $\hat{T}_1$ and $\hat{T}_2$ be given by
\begin{align}
\tilde{T}_1&:=\ts\frac{1}{\gamma} \med(|y_1\ccirc|,\dots,|y_{n_1}\ccirc|)\label{tilde1}\\[3pt]
\tilde{T}_2^2&:= \ts\frac{1}{\gamma^2 n_2} \Big((y_{n_1+1}\ccirc)^2+\cdots + (y_{n_1+n_2}\ccirc)^2\Big).\label{tilde2}
\end{align}
It is convenient to work in terms of these variables, since their limiting distributions may be computed exactly. Due to the fact that $\ts\frac{y_1\ccirc}{\gamma \|x\|_1},\dots,\ts\frac{y_{n_1}\ccirc}{\gamma \|x\|_1}$ is an i.i.d. sample from the standard Cauchy distribution $C(0,1)$, the asymptotic normality of the sample median implies
\begin{equation}\label{clt1}
\sqrt{n/2}\big(\ts\frac{\tilde{T}_1}{\|x\|_1}-1\big)\overset{\mathcal{L}}{\to} N(0,\tau_1^2),
\end{equation}
where $\tau_1^2=\pi^2/8$. Additional details may be found in~\cite[Theorem 9.2]{davidorder} and~\cite[Lemma 3]{pingliJMLR}. Similarly, the variables $(\ts\frac{y_{n_1+1}\ccirc}{\gamma \|x\|_2})^2,\dots,(\ts\frac{y_{n_1+n_2}\ccirc}{\gamma \|x\|_2})^2$ are an i.i.d. sample from the chi-square distribution on one degree of freedom, and so it follows from the delta method that
\begin{equation}\label{clt2}
\sqrt{n/2}\big(\ts\frac{\tilde{T}_2}{\|x\|_2}-1\big)\overset{\mathcal{L}}{\to} N(0,\tau_2^2),
\end{equation}
where $\tau_2^2=1/2$. Note that in proving the last two limit statements, we intentionally scaled the variables $y_i\ccirc$ in such a way that their distributions did not depend on any model parameters.  It is for this reason that the limits hold even when the model parameters are allowed to depend on $n$. We conclude from the limits~\eqref{clt1} and~\eqref{clt2} that for any $\alpha\in(0,1/2)$,
\begin{equation}\label{lim1}
\P\Big( \ts\frac{\tilde{T}_1}{\|x\|_1} \in [1-\ts\frac{\tau_1 z_{1-\alpha}}{\sqrt{n/2}}, 1+\ts\frac{\tau_1 z_{1-\alpha}}{\sqrt{n/2}}]\Big) = 1-2\alpha +o(1),
\end{equation}
and
\begin{equation}\label{lim2}
\P\Big( \ts\frac{\tilde{T}_2}{\|x\|_2} \in [1-\ts\frac{\tau_2 z_{1-\alpha}}{\sqrt{n/2}}, 1+\ts\frac{\tau_2 z_{1-\alpha}}{\sqrt{n/2}}]\Big) = 1-2\alpha +o(1).
\end{equation}

~\\
\noindent We now relate $\hat{T}_1$ and $\hat{T}_2$ in terms of intervals defined by $\tilde{T}_1$ and $\tilde{T}_2$. Since the noise variables are bounded by $|\e_i|\leq \sigma_0$, and $y_i=y_i\ccirc+\e_i$,  it is easy to see that 
$$\hat{T}_1\in [\tilde{T}_1-\ts\frac{\sigma_0}{\gamma},\tilde{T}_1+\ts\frac{\sigma_0}{\gamma}].$$
Consequently, if we note that $\ts\frac{\sigma_0}{\gamma \|x\|_1}\leq\ts\frac{\sigma_0}{\gamma \|x\|_2}=\rho$, then we may write
\begin{equation}\label{interval1} 
\ts\frac{\hat{T}_1}{\|x\|_1} \in \Big[ \ts\frac{\tilde{T}_1}{\|x\|_1} -\rho, \ \ts\frac{\tilde{T}_1}{\|x\|_1}+\rho\Big].
\end{equation}
To derive a similar relationship involving $\hat{T}_2$ and $\tilde{T}_2$, if we write $\hat{T}_2$ in terms of $\|(y_{n_1},\dots,y_{n_1+n_2})\|_2$ and apply the triangle inequality, it follows that
\begin{equation}\label{interval2}
\ts\frac{\hat{T}_2}{\|x\|_2} \in \Big[ \ts\frac{\tilde{T}_2}{\|x\|_2} -\rho, \ \ts\frac{\tilde{T}_2}{\|x\|_2}+\rho\Big].
\end{equation}
The proof may now be completed by assembling the last several items. Recall the parameters $\delta_n$ and $\eta_n$, which are given by
\begin{align}
\delta_n &= \delta_n(\alpha,\rho)= \ts\frac{\tau_1 z_{1-\alpha}}{\sqrt{n/2}}+\rho\\[5pt]
\eta_n &= \eta_n(\alpha,\rho)= \ts\frac{\tau_2 z_{1-\alpha}}{\sqrt{n/2}}+\rho.
\end{align}
Combining the limits~\eqref{lim1} and~\eqref{lim2} with the intervals~\eqref{interval1} and~\eqref{interval2}, we have the following asymptotic bounds for the statistics $\hat{T}_1$ and $\hat{T}_2$,
\begin{equation}\label{lim1again}
\P\Big( \ts\frac{\hat{T}_1}{\|x\|_1} \in [1-\delta_n, 1+\delta_n]\Big) \geq 1-2\alpha +o(1),
\end{equation}
and
\begin{equation}\label{lim2again}
\P\Big( \ts\frac{\hat{T}_2}{\|x\|_2} \in [1-\eta_n, 1+\eta_n]\Big) \geq 1-2\alpha +o(1).
\end{equation}
Due to the independence of $\hat{T}_1$ and $\hat{T}_2$, and the relation
$$\ts\sqrt{\frac{\hat{s}(x)}{s(x)}}=\frac{\hat{T}_1/\|x\|_1}{\hat{T}_2/\|x\|_2},$$
we conclude that
\begin{equation}\label{noisePropUpper}
\P\Big(\ts\sqrt{\frac{\hat{s}(x)}{s(x)}} \in \big[\frac{1-\delta_n}{1+\eta_n},\frac{1+\delta_n}{1-\eta_n}\big]\Big) \geq (1-2\alpha)^2+o(1). 
\end{equation}
\qed
\subsection{Proof of Theorem 2.}
The proof of Theorem 2 is almost the same as the proof of Theorem 1 and we omit the details. One point of difference is that in Theorem 1, the bounding probability is $(1-2\alpha)^2$, whereas in Theorem 2 it is $(1-2\alpha)$. The reason is that in the case of Theorem 2, the condition $\breve{T}_1/\|x\|_1 \in [1-\varrho,1+\varrho]$ holds with probability 1, whereas the analogous statement $\hat{T}_1/\|x\|_1 \in [1-\rho,1+\rho]$ holds with probability $1-2\alpha$ in the case of Theorem 1.

\subsection{Proof of Theorem 3.}

The following lemma illustrates the essential reason why estimating $s(x)$ is difficult in the deterministic case. The idea is that  for any measurement matrix $A$,  it is possible to find two signals that are indistinguishable with respect to $A$, and yet have very different sparsity levels in terms of $s(\cdot)$. We prove Theorem 3 after giving the proof of the lemma.

\begin{lemma}\label{impslem}
Let $A\in \R^{n\times p}$ be an arbitrary matrix, and let $x\in\R^p$ be an arbitrary signal. Then, there exists a non-zero vector $\tilde{x}\in\R^p$ satisfying $Ax=A\tilde{x}$, and
\begin{equation}\label{claim2}
s(\tilde{x})\geq \frac{p-n}{(1+2\sqrt{2\log(2p)})^2}. 
\end{equation}
\end{lemma}

\noindent \emph{Proof of Lemma 1.} By H\"older's inequality, 
$\|\tilde{x}\|_1/\|\tilde{x}\|_2\geq \|\tilde{x}\|_2/ \|\tilde{x}\|_{\infty},$
and so it suffices to lower-bound the second ratio. The overall approach to finding a dense vector $\tilde{x}$ is to use the probabilistic 
method. Let $B\in \R^{p\times(p-r)}$ be a matrix 
whose columns are an orthonormal basis for the null space of $A$, where 
$r=\text{rank}(A)$. Also define the scaled matrix $\tilde{B}:=\|x\|_{\infty}B$.
Letting $z\in \R^{p-r}$ be a standard Gaussian vector, we will consider 
$\tilde{x} := x+\tilde{B}z$, which satisfies $Ax=A\tilde{x}$ for all realizations 
of $z$. We begin the argument by defining the function
\begin{equation}\label{func}
\textstyle
f(z):= \|x+\tilde{B}z\|_2 - c(n,p)\cdot\|x+\tilde{B}z\|_{\infty},
\end{equation}
where 
$$c(n,p):= \frac{\sqrt{p-n}}{1+2\sqrt{2\log(2p)}}.$$
The proof amounts to showing that the event $\{f(z)>0\}$ holds with 
positive probability. To see this, notice that the event $\{f(z)>0\}$ is equivalent to 
$$\frac{\|\tilde{x}\|_2}{\, \, \|\tilde{x}\|_{\infty}}=\frac{\|x+\tilde{B}z\|_2}{\: \: \|x+\tilde{B}z\|_{\infty}} > \frac{\sqrt{p-n}}{1+2\sqrt{2\log(2p)}}. $$
We will prove that $\P(f(z)>0)$ is positive by showing that $\E[f(z)]>0$, and 
this will be accomplished by lower-bounding the expected value of 
$\|x+\tilde{B}z\|_2$, and upper-bounding the expected value of 
$\|x+\tilde{B}z\|_{\infty}$.

 First, to lower-bound $\|x+\tilde{B}z\|_2,$ we begin by considering 
 the variance of $\|x+\tilde{B}z\|_2$, and use the fact that 
 $\|\tilde{B}z\|_2^2= z\ttop \tilde{B}\ttop \tilde{B}z = \|x\|_{\infty}^2 \, \|z\|_2^2$, obtaining
\begin{equation}\label{variance}
\begin{split}
\E\|x+\tilde{B}z\|_2 &= \sqrt{\E \|x+\tilde{B}z\|_2^2 - \var \|x+\tilde{B}z\|_2}\\
&=\sqrt{\|x\|_2^2+\|x\|_{\infty}^2\,(p-r)-\var \|x+\tilde{B}z\|_2}.
\end{split}
\end{equation} 
To upper-bound the variance, we use the Poincar\'e inequality for the 
standard Gaussian measure on $\R^{p-r}$~\cite{Beckner}. Since the 
function $g(z):=\|x+\tilde{B}z\|_2$ has a Lipschitz constant equal to 
$\|\tilde{B}\|\op = \|x\|_{\infty}$ with respect to the Euclidean norm, it 
follows that $\|\nabla g(z)\|_2 \leq \|x\|_{\infty}$.
Consequently, the Poincar\'e inequality implies
 $$\var \|x+\tilde{B}z\|_2\leq \|x\|_{\infty}^2.$$
Using this in conjunction with the inequality \eqref{variance}, and the fact 
that $r=\text{rank}(A)$ is at most $n$, we obtain the lower bound
\begin{equation}\label{2norm}
\E\|x+Bz\|_2 \geq \sqrt{\|x\|_2^2+ \|x\|_{\infty}^2(p-n)-\|x\|_{\infty}^2}.
\end{equation}

The second main portion of the proof is to upper-bound $\E\|x+\tilde{B}z\|_{\infty}$. 
Since $\|x+\tilde{B}z\|_{\infty}\leq \|x\|_{\infty}+\|\tilde{B}z\|_{\infty}$, it is enough 
to upper-bound $\E\|\tilde{B}z\|_{\infty}$, and we will do this using a version of 
Slepian's inequality. If $\tilde{b}_i$ denotes the $i^{\text{th}}$ row of $\tilde{B}$, 
define $g_i=\langle \tilde{b}_i,z\rangle$, and let $w_1,\dots, w_p$ be i.i.d. $N(0,1)$ 
variables. The idea is to compare the Gaussian process $g_i$ with the Gaussian 
process $\|x\|_{\infty}w_i$. By Proposition A.2.6 in van der Vaart and 
Wellner~\cite{vaartWellner}, the inequality
$$\E\|\tilde{B}z\|_{\infty}  = \E\left[ \max_{i=1,\dots, p}|g_i| \right]\leq 2\|x\|_{\infty}\,\E \left[\max_{i=1,\dots,p}|w_i|\right],$$
holds as long as the condition $\E(g_i-g_j)^2\leq \|x\|_{\infty}^2\, \E(w_i-w_j)^2$ 
is satisfied for all $i,j\in \{1,\dots,p\}$, and this is simple to verify.
To finish the proof, we make use of a standard bound for the expectation of Gaussian maxima 
$$
\E \left[\max_{i=1,\dots,p}|w_i|\right] < \sqrt{2\log(2p)},
$$
which follows from a modification of the proof of Massart's finite class 
lemma~\cite[Lemma 5.2]{massartFiniteClass}\footnote{The ``extra'' factor 
of 2 inside the logarithm arises from taking the absolute value of the $w_i$.}.
Combining the last two steps, we obtain
\begin{equation}\label{inftynorm}
\E\|x+Bz\|_{\infty}< \|x\|_{\infty} + 2\|x\|_{\infty}\sqrt{2\log(2p)}.
\end{equation}
Finally, applying the bounds \eqref{2norm} and \eqref{inftynorm} 
to the definition of the function $f$ in $\eqref{func}$, we have 
\begin{equation}
\begin{split}
\frac{\E\|x+Bz\|_2 }{\ \E\|x+Bz\|_{\infty}}&>\frac{\sqrt{\|x\|_2^2+ \|x\|_{\infty}^2(p-n)-\|x\|_{\infty}^2}}{\|x\|_{\infty} + 2\|x\|_{\infty}\sqrt{2\log(2p)}}\\
~\\
&=\frac{\sqrt{\frac{\|x\|_2^2}{\:\: \|x|_{\infty}^2}+ (p-n)-1}}{1 + 2\sqrt{2\log(2p)}}\\
~\\
&\geq \frac{\sqrt{p-n}}{1 + 2\sqrt{2\log(2p)}},
\end{split}
\end{equation}
which proves $\E[f(z)]>0$, as needed. \qed 
~\\

We now apply Lemma 3 to prove Theorem 3.\\

\noindent \emph{Proof of Theorem 3.} We begin by making several reductions. First, it is enough to show that 
\begin{equation}\label{red1}
\inf_{A\in \R^{n\times p}} \inf_{\delta:\R^n\to \R} \:\sup_{x\in\R^p\setminus\{0\}}\Big|\delta(Ax)-s(x)\Big|\geq  \frac{p-n-1}{2(1+2\sqrt{2\log(2p)})^2}.
 \end{equation}
To see this, note that the general inequality $s(x)\leq p$ implies
$$\big|\ts\frac{\delta(Ax)}{s(x)}-1\big| \geq \ts\frac{1}{p}\big|\delta(Ax)-s(x)\big|,$$
and we can optimize over both sides with $p$ being a constant. Next, for any fixed matrix $A\in \R^{n\times p}$, it is enough to show that
\begin{equation}\label{red1}
\inf_{\delta:\R^n\to \R} \:\sup_{x\in\R^p\setminus\{0\}}\Big|\delta(Ax)-s(x)\Big|\geq  \frac{p-n-1}{2(1+2\sqrt{2\log(2p)})^2},
 \end{equation} 
as we may take the infimum over all matrices $A$ without affecting the right hand side. To make a third reduction, it is enough to prove the same bound when $\R^p\setminus\{0\}$ is replaced with any smaller set, as this can only make the supremum smaller. In particular, we will replace $\R^p\setminus\{0\}$ with a two-point subset   $\{x^{\circ},\tilde{x}\}\subset \R^p\setminus\{0\}$, where by Lemma 1, we may choose $\tilde{x}$ and $x^{\circ}$ to satisfy $Ax^{\circ}=A\tilde{x}$, as well as 
$$s(x^{\circ})=1, \text{  \  and \ \ } s(\tilde{x})\geq \frac{p-n}{2(1+2\sqrt{2\log(2p)})^2}.$$
We now aim to prove that
\begin{equation}\label{red2}
\inf_{\delta:\R^n\to \R} \:\sup_{x\in\{x^{\circ},\tilde{x}\}}\Big|\delta(Ax)-s(x)\Big|\geq  \frac{p-n-1}{2(1+2\sqrt{2\log(2p)})^2},
 \end{equation} 
 and we will accomplish this using the classical technique of constructing a Bayes procedure with constant risk. For any decision rule $\delta:\R^n\to \R$ and any point \mbox{$x\in \{x^{\circ},\tilde{x}\}$,} define the (deterministic) risk function 
 $$R(x, \delta):=\Big| \delta(Ax)-s(x)\Big|.$$
 Also, for any prior $\pi$ on $\{x^{\circ},\tilde{x}\}$, define
 $$r(\pi,\delta):= \int R(x,\delta) d\pi(x).$$ 
  By Propositions 3.3.1 and 3.3.2 of~\cite{bickelDoksum}, the inequality~\eqref{red2} holds if there exists a prior distribution $\pi^*$ on $\{x^{\circ},\tilde{x}\}$ and a decision rule $\delta^*:\R^n\to\R$ with the following three properties:
\begin{enumerate}
\item The rule $\delta^*$ is Bayes for $\pi^*$, i.e. $r(\pi^*,\delta^*)= \inf_{\delta} r(\pi^*,\delta)$.
\item The rule $\delta^*$ has constant risk over $\{x^{\circ},\tilde{x}\}$, i.e. $R(x^{\circ},\delta^*)= R(\tilde{x},\delta^*)$.
\item The constant value of the risk of $\delta^*$ is at least $ \frac{p-n-1}{2(1+2\sqrt{2\log(2p)})^2}$.
\end{enumerate} 
To exhibit $\pi^*$ and $\delta^*$ with these properties, we define $\pi^*$ to be the two-point prior that puts equal mass at $x^{\circ}$ and $\tilde{x}$, and we define $\delta^*$ to be the trivial decision rule that always returns the average of the two possibilities, namely  $\delta^*(Ax)=\frac{1}{2}(s(\tilde{x})+s(x^{\circ}))$.  It is simple to check the second and third properties, namely that $\delta^*$ has constant risk equal to $\frac{1}{2}|s(\tilde{x})-s(x^{\circ})|$, and that this risk is at least $ \frac{p-n-1}{2(1+2\sqrt{2\log(2p)})^2}$. It remains to check that $\delta^*$ is Bayes for $\pi^*$. This follows easily from the triangle inequality, since for any $\delta$,
\begin{equation}
\begin{split}
r(\pi^*,\delta)&=\ts\frac{1}{2}\Big|\delta(Ax)-s(\tilde{x})\Big|+\frac{1}{2}\Big|\delta(Ax)-s(x^{\circ})\Big|,\\
&\geq\ts \frac{1}{2}\Big|s(\tilde{x})-s(x^{\circ})\Big|\\
&=r(\pi^*,\delta^*).
\end{split}
\end{equation}
\qed

{\bibliography{CauchyBib.bib}}

\bibliographystyle{plainnat}

\end{document}